\documentclass[twocolumn]{aastex631}
\usepackage{xspace} % intelligent space: space when there is word next, no-space when followed by a period

\usepackage{graphicx}	% Including figure files
\usepackage{amsmath}	% Advanced maths commands
\usepackage{amssymb}	% Extra maths symbols
\usepackage{hyperref}
\newcommand{\heii}{He\,\textsc{ii}\xspace}
\newcommand{\heiioptic}{He\,\textsc{ii}~$\lambda4686$\xspace}
\newcommand{\heiiUV}{He\,\textsc{ii}~$\lambda1640$\xspace}
\newcommand{\hbeta}{\ensuremath{\mathrm{H{\beta}}}\xspace}
\newcommand{\Msun}{\ensuremath{\mathrm{M}_{\odot}}\xspace}
\newcommand{\Zsun}{\ensuremath{\mathrm{Z}_{\odot}}\xspace}

\begin{document}

\title{The Significant Contribution of Supersoft X-ray Sources to Nebular HeII Line Emission}

\correspondingauthor{Dian P. Triani}
\email{pipit.triani@cfa.harvard.edu}

\author[0000-0002-4752-128X]{Dian P. Triani}
\affiliation{Institute for Theory and Computation, Harvard-Smithsonian Center for Astrophysics \\
Cambridge, MA 02138, USA}
\affiliation{Center for Astrophysics \text{\textbar} Harvard \& Smithsonian, 60 Garden Street, Cambridge, MA 02138, USA}

\author[0000-0003-0972-1376]{Rosanne Di Stefano}
\affiliation{Institute for Theory and Computation, Harvard-Smithsonian Center for Astrophysics \\
Cambridge, MA 02138, USA}
\affiliation{Center for Astrophysics \text{\textbar} Harvard \& Smithsonian, 60 Garden Street, Cambridge, MA 02138, USA}

\author[0000-0003-4512-8705]{Tiger Yu-Yang Hsiao}
\affiliation{Center for Astrophysics \text{\textbar} Harvard \& Smithsonian, 60 Garden Street, Cambridge, MA 02138, USA}
\affiliation{Center for Astrophysical Sciences, Department of Physics and Astronomy, The Johns Hopkins University, 3400 N Charles St. Baltimore, MD 21218, USA}
\affiliation{Space Telescope Science Institute (STScI), 3700 San Martin Drive, Baltimore, MD 21218, USA}

\author[0000-0001-8152-3943]{Lisa J. Kewley}
\affiliation{Institute for Theory and Computation, Harvard-Smithsonian Center for Astrophysics \\
Cambridge, MA 02138, USA}
\affiliation{Center for Astrophysics \text{\textbar} Harvard \& Smithsonian, 60 Garden Street, Cambridge, MA 02138, USA}



\begin{abstract}

Nebular spectral lines provide insight into the properties of the interstellar medium (ISM) and the ionizing radiation within galaxies. The presence of high-energy ionization lines such as \heii indicates the existence of ionizing photons with energies exceeding the second ionization energy of helium ($54 \mathrm{eV})$. There is an enigma surrounding the origin of these lines observed in star-forming galaxies because stellar ionization cannot account for such high energy emission. This paper proposes that supersoft X-ray sources (SSSs) may produce the \heii ionization lines in star-forming galaxies. We model the spectra of SSSs using blackbody radiation and add them to the young stellar population spectra to represent the overall spectra of galaxies. Using a photoionization model, we predict the resulting \heiioptic and \hbeta line fluxes and inspect the contribution of SSSs to the elevation of the \heiioptic/\hbeta ratio in star-forming galaxies, both at low and high redshifts. We find that incorporating a blackbody with temperatures between $kT = 10-100 \mathrm{eV}$ can boost the \heiioptic/\hbeta line ratio to the levels observed in local galaxies by SDSS and in early galaxies by NIRSpec. This blackbody temperature range aligns with the observed temperatures of SSSs. The number of SSSs in spiral galaxies listed in Chandra catalogues, and our estimates of the total population, confirms that SSSs are promising candidates for the source of the \heii ionization.

\end{abstract}


%Need a ref. for WR stars.
%In the last sentence of paragraph 2 in the introduction, do we mean "in the rest-frame ultraviolet"?

\section{Introduction} \label{sec:intro}

%During a period known as the epoch of reionization, the neutral gas in the Universe is ionized by the first stars in the first galaxies until it becomes fully ionized as observed today \citep{Barkana16}. This period is crucial because it heavily impacts subsequent galaxy evolution. Ionizing photons heat up the intergalactic medium \citep{Fan06}, altering the gas infall into galaxies, which later influences the cooling and star-formation processes. Understanding cosmic reionization is a crucial quest in galaxy evolution. To do so, we need to examine the gas properties and radiation field of distant galaxies.

Nebular lines arise from the ionization of neutral gas by a radiation field. Hence, such lines provide strong diagnostics for both the properties of the gas in the interstellar medium (ISM) and the ionizing radiation source. Combinations of the ratios of rest frame UV and optical line ratios have been used to infer gas-phase metallicity, electron temperature, and ionization parameters \citep[see][for a review]{Kewley19}. It has also been widely used to distinguish among primary ionization sources, such as star formation, active galactic nuclei (AGN), and other energetic sources \citep{Baldwin81, Kewley01, Kauffmann03}. A common tool for this classification is the BPT diagram \citep{Baldwin81}, which uses ratios of strong emission lines (e.g., [OIII]/H$\beta$ and [NII]/H$\alpha$) to separate star-forming regions from AGN and other ionization mechanisms in a diagnostic diagram.

The presence of spectral lines with different ionization potentials is closely related to the shape of the radiation field powering the emission. High-ionization spectral features, such as \heiioptic, arise from the electron transition between the n=4 and n=3 energy levels of singly ionized helium ($\mathrm{He^+}$) and require ionizing photons with energies above 54 eV. In this work, we use \heiioptic as the default reference for \heii unless otherwise specified, in which case we refer to \heiiUV. These high-energy photons are typically produced by energetic sources, such as active galactic nuclei (AGN). Galaxies hosting AGN commonly exhibit multiple high-ionization lines and occupy the top-right region of the BPT diagram \citep{Baldwin81}. Conversely, galaxies lacking AGN exhibit weaker ionization and fall in the bottom-left region, where stellar sources are thought to be the dominant ionization mechanism \citep{Kewley01, Kauffmann03}.

Although stellar emission typically cannot produce high-energy photons, high-ionization lines like \heiioptic have been detected in local metal-poor star-forming galaxies with no AGN features \citep{Garnett91, Shirazi12, Senchyna17, Senchyna19, Berg19}. This observation is often attributed to the harder radiation fields of low-metallicity stars, as reduced metal content increases the temperature of massive stars, leading to more energetic ionizing photons \citep{Stanway19}. However, the temperature of metal-poor stars is not sufficient to produce \heii lines. 

Normal stars generally do not emit photons with energies sufficient to ionize \heii, except for exotic species such as Wolf-Rayet (WR) stars, which can reach temperatures high enough to produce such photons \citep{Schaerer96}. However, WR stars are short-lived and rare in low-metallicity stellar populations \citep{Crowther07}. In theory, Population III (PopIII) stars, which are nearly metal-free, are predicted to have hard spectra capable of producing \heii-ionizing photons \citep{Cassata13, Oh01}. However, the direct detection of such systems remains elusive.

Furthermore, several high-ionization lines have been detected in the rest-frame UV spectra of $z>6$ galaxies \citep{Stark15,Schmidt17}. The presence of these lines suggests strong ionizing radiation fields in early galaxies, highlighting the importance of understanding the ionization mechanisms that drive such emission. Addressing this problem is key to unraveling the heating and ionization processes at high redshift, including those that governed the reionization era.

Several mechanisms have been proposed to address this problem. High-mass X-ray binaries (HMXBs) and ultraluminous X-ray sources (ULXs) have been suggested as potential producers of \heii-ionizing photons due to their high X-ray luminosities ($L = 10^{39} \mathrm{erg/s}$) in the 0.3–10 keV energy range \citep{Simmonds21, Schaerer19, Senchyna20}. However, \citet{Saxena20} found no correlation between the X-ray luminosity and the presence of the $\mathrm{He_{II}}\lambda1640$ line in $z \approx 2.3 -3.6$ galaxies from VANDELS survey, making it unlikely that X-ray binaries are the primary ionization source of \heii. A similar conclusion was reached by \citet{Kehrig21}, who studied the metal-poor galaxy IZw18 and showed that the model using HMXBs fails to reproduce the observed \heii ionization. These findings suggest that other mechanisms must contribute significantly to \heii production in these environments.

Soft X-rays (0.1–2.0 keV), in general, are efficient at producing \heii ionization. \citet{Olivier22} found that adding a blackbody component to a photoionization model improves \heii emission and other high-ionization lines. The origin of this blackbody radiation has been discussed in several works. For example, \citet{Chen15} proposed accreting white dwarfs as the primary producers of soft X-rays, but their results show an elevated \heiioptic/\hbeta ratio only in elliptical galaxies with inactive star formation. Similar results were presented by \citet{Garofali24}, where \heii enhancements were observed only in older stellar populations, not in star-forming galaxies. \citet{Oskinova22} demonstrated that superbubbles can produce ionizing radiation when the plasma temperature is sufficiently low, while \citet{Sarkar22} discussed excess soft X-ray radiation in galactic winds. Additional mechanisms such as radiative shocks \citep{Thuan05, Izotov12, Plat19}, photon leakage \citep{Perez-Montero20}, and binary-produced processes \citep{Stanway19} have also been considered. However, none of these mechanisms alone appear sufficient to explain the observed \heiioptic emission in star-forming galaxies, where the ionizing conditions remain poorly understood.

A critical ionizing source that has not been extensively explored in the context of \heiioptic ionization is supersoft X-ray sources (SSSs). SSSs emit predominantly in the 10–100 eV energy range, aligning closely with the ionization potential required for \heii (54.4 eV), unlike XRBs that emit higher-energy photons ($>0.3$ keV) inefficient for \heii ionization \citep{Garofali24, Oskinova22, Kehrig21}. One of the first and most well-studied SSSs, CAL 83 in the Large Magellanic Cloud, is surrounded by a large nebula exhibiting strong \heiioptic emission, directly demonstrating the ionizing potential of these sources \citep{Crampton87, Schmidtke04}. With typical luminosities of $10^{36} - 10^{39}$~erg~s$^{-1}$ \citep{Kahabka97, Greiner96, Kahabka08}, SSSs provide a highly targeted ionizing spectrum that fills the gap left by other mechanisms. Observations show that SSSs are present in environments ranging from nearby galaxies, such as M31 and M33, to more distant star-forming systems \citep{DiStefano03, Orio10, Galiullin21}. 

In this paper, we propose that SSSs are key contributors to \heiioptic production in star-forming galaxies, particularly in cases where other sources fail to fully explain the observed flux. We focus on the optical \heiioptic emission, which is more commonly used in optical spectroscopic surveys, allowing for direct comparison with observed HeII fluxes in star-forming galaxies. By incorporating SSSs into a photoionization model, this study aims to provide a more complete understanding of the ionizing photon budget and address long-standing challenges in explaining \heiioptic emission.

We expand upon the characteristics and possible natures of SSSs in \S \ref{sec:SSSs}. 
Because the definition of SSSs is based on phenomenology, we do not expect that they represent sources with a single physical nature. Their contribution to the ionization of the gas in galaxies is, however, independent of their fundamental natures. It depends only on the numbers of SSSs populating the galaxy, and the distribution of luminosities and temperatures among the SSSs.

To model SSSs as the source of high ionization lines, we take a blackbody spectra with various temperatures and luminosities as inputs to the MAPPINGS photoionization code \citep{Sutherland18}. The resulting optical lines are then compared to observations of \heii emission in star-forming galaxies. For a more realistic approach in modelling the galaxy spectra with SSSs, we combine the blackbody spectra with young stellar populations from BPASS with various scalings for the blackbody contribution to the total spectra. We found that a substantial, but likely realistic, contribution of blackbody spectra is required to produce the emission line fluxes observed in star forming galaxies. The methodology is described in \S \ref{sec:modeling} and we introduced our observation sample in \S \ref{ssec:observation}. The results are presented in \S \ref{sec:results} and we discuss the implication of SSSs to \heii ionization in \S \ref{sec:discussion}. Our conclusions are summarised in \S \ref{sec:summary}.

\section{Why Supersoft X-ray Sources?} \label{sec:SSSs}

\subsection{Classical SSSs}

The class of luminous supersoft X-ray sources is defined phenomenologically, in terms of the temperatures and luminosities of the sources. The first-discovered SSSs were in the Magellanic Clouds \citep{LongHelfandGrabelsky}. During the early 1990s, the ROSAT observatory established the class of SSSs by discovering additional Magellanic Cloud SSSs, as well as M31 SSSs and a small number in the Galaxy \citep{GreinerSSS}. These  $\sim 40$ SSSs had values  of $k\, T$ in the range from about $30$~eV to roughly $90$~eV, and luminosities in the range a few $\times 10^{35}$~erg~s$^{-1}$ to a few $\times 10^{38}$~erg~s$^{-1}$.  
%1981ApJ...248..925L

Sources in the temperature range of SSSs can be highly effective providers of the photons needed to create high-energy ionization states. The magnitude of their influence as ionizers within and around the galaxies they populate depends on the sizes of SSS populations within galaxies. 
Establishing the size of a galaxy's SSS population is challenging because a large fraction of the radiation they emit is absorbed before reaching our detectors. Thus, the very circumstance that makes them important ionization sources decreases the detectability of SSSs.   We can therefore detect only a fraction of the SSSs present, especially in galaxies containing gas and dust.

\citet{DiStefano94} used the ROSAT-observed SSSs to estimate the size of the underlying populations in the Magellanic Clouds, M31, and in the Galaxy. They constructed several models for the distributions of gas and dust in each of these systems, and then seeded each galaxy with sources having the properties of the observed SSSs. The range of galaxy models allowed an estimate of the range of population sizes.
It was found that in the Magellanic Clouds, the $4$ ROSAT-observed SSSs could represent a total population of similar SSSs as small as 22 and as large as $100$.  (These are sources that would be ``on'', but not detectable because of absorption.)
The total SSS population numbers in the Milky Way range from 400 to 2900, with a most likely value of $1000$. 
Because the numbers of SSSs reported in M31 was large ($25$), and the distance is greater, thereby reducing the count rate, the estimated population of SSSs was larger: $800-5000$ with a likely value of  $2500$. 
While the ranges of population sizes consistent with the early data were large, the common factor is that ROSAT was detecting a small fraction of the SSSs active at any given time. This fraction ranges from on the order of $10\%$
in the Magellanic Clouds to about a percent in M31. Furthermore, the fraction of the SSSs detected is highly sensitive to the source's temperature, increasing with increasing temperature. 

If we take $10^{37}$~erg~s$^{-1}$ to be a typical SSS luminosity, the Milky Way and M31 would have minimum SSS total galactic 
luminosities (i.e. using the lowest possible estimated fractions) of $4 \times 10^{39}~$erg~s$^{-1}$ and $8 \times 10^{39}~$erg~s$^{-1}$, respectively. 
Note that the work described here makes no assumptions about the nature or natures of the SSSs.  

\subsection{Possible Natures of SSSs}
\subsubsection{Hot White Dwarfs}
While the population estimates discussed above are independent of the physical natures of SSSs, the galactic environments they inhabit, does however, depend on the natures of the sources. 

The ranges of luminosities and temperatures of the first-discovered “classical” SSSs, are consistent with the effective radii of hot white dwarfs (WDs). There is evidence that many of the Local Group SSSs are WDs. The largest number are post-nova WDs. 
A nova explosion occurs when matter accreted from a stellar companion reaches a critical mass, whose value depends on the WD mass, the accretion rate, and the composition of the accreting matter \citep{Bode08}.  
 After the ejecta have become diffuse enough to allow soft X-ray photons to escape, the nova will be detected as an SSS if residual nuclear burning is occurring or if the
WD has not yet cooled. A program of X-ray observations of optical novae in M31 has found the largest number of SSSs in the Local Group \citep{Henze2014}. %Note that the cooling of the WD may transform it into an HSS. Other SSSs have been observed as the central stars in planetary nebulae.

Another type of system observed as SSSs are close binaries with orbital periods on the order of 10 hours to about a day.  These are hypothesized to be WDs that are hot because they are burning incoming matter donated by a companion \citep{vdH1992}. Because nuclear burning requires high accretion rates, these are special binaries, often referred to as  “close binary supersoft sources (CBSSs) \citep{diStefano96}. Perhaps a better name for them would be close binary nuclear burning white dwarfs, since not all of the systems identified as CBSSs are always observed as supersoft \citep{Greiner96, Starrfield04}. There may even be a subset of the class that is rarely ever detected as an SSS. 

Symbiotic stars in which a wide orbit companion donates mass to a WD at high rates, generally through winds or through a combination of winds and Roche-lobe filling, are also observed as SSSs \citep{Jordan94, Sokoloski06}. Some symbiotics have rates of mass transfer that are too low to permit quasisteady nuclear burning but are high enough to lead to recurrent novae. A well-known example is RS Ophiuchi, a symbiotic recurrent nova, where the WD periodically undergoes thermonuclear runaways, followed by a supersoft X-ray phase \citep{Osborne11}. Another remarkable case is a nova in M31 identified by \citet{Henze15} which erupts almost annually and has been observed as a supersoft source.
A key significance of nuclear-burning WDs in CBSSs is their ability to retain accreted matter, which may lead to accretion-induced collapse into a neutron star \citep{vdH1992} or, in some cases, a Type Ia supernova \citep{Rappaport1994}

\subsubsection{Neutron Stars and Black Holes}
At least one of the ``classical'' SSSs has been hypothesized to be an accreting black hole (BH) \citep{CowleyBH}.  Whether or not that system contains a BH, there is reason to believe that some SSSs in external galaxies contain BHs.
The likelihood that a subset of SSSs have BH or neutron star (NS) accretors is suggested by extensions of the class of SSSs that have been discovered in other galaxies. As we will discuss below, some SSSs have luminosities greater than $10^{39}$~erg~s$^{-1}$. This is too large a luminosity to be produced by WDs, see section \ref{ssec:extension}

\subsection{SSS Characteristics Derived from Galactic Studies}

\subsubsection{Duty Cycle}
Some of the known CBSSs have been observed in X-ray ``off'' states, during which there is no X-ray emission. Transitions between SSS states and X-ray ``off'' states can take place over intervals of days \citep{Greiner02}. During `off'' states, the optical emission, most of which is provided by the accretion disk, increases by about a magnitude. This is consistent with the X-ray emitting region having a larger effective radius, thereby providing more EUV and UV irradiation of the accretion disk. This combination of X-ray and optical behavior suggests that the photosphere of the SSS has expanded, moving the high-energy tail out of the X-ray regime toward longer-wavelength radiation \citep{Greiner02}. The known CBSSs spend enough of the time with SSS temperatures to provide a good chance of X-ray detection. But there may be a subset of sources with the same physical nature that have temperatures above $10-20$~keV rarely if at all.

\subsubsection{Nebulae}
Given their spectrum, SSSs are capable of exciting high-energy ionization states. The photons that travel from the source through the ISM have the ability to ionize many solar masses, so that regions
 of $\sim 10$ pc around each source could be ionized \citep{Rappaport1994}. CAL~83 has such a nebula \citep{RemillardNeb}. Other SSSs, however, have not been seen to have them. One possible reason is that the duty cycle of the SSS phase of binary evolution is too short. Simple calculations indicate that the duty cycle would have to be on the order of a few percent in order to explain the lack of SSSs \citep{ChiangNeb}. The other possibility is that the ISM in regions around most SSSs is not as dense as is the region around CAL83 \citep{WoodsGilNeb}. This could be because of the position of the sources within the galaxy. But it is also feasible that the sources themselves lower the local density. Jets, seen in some SSSs are one way in which the ISM can be pushed away from the source. In addition, if a nebula forms it is expected to expand. The radiation from SSSs in regions of moderate to low density would escape and ionize gas far from the SSS. 

\subsection{Extensions of the Class of SSSs Derived from Extra Galactic Studies}
\label{ssec:extension}
The calculations of \citet{DiStefano94} used the then-known SSSs as input, assuming that they were typical representatives of the class. With the advent of {\sl Chandra}  and {\sl XMM-Newton}, it became possible to detect SSSs in other galaxies.  Typical galaxy populations of SSSs are on the order of a dozen per galaxy \citep{Pence01, Swartz02, DiStefano03, DiStefano03a}. \citet{DiStefano94} showed that, as expected, the effects of increased absorption are to make it more difficult to observe the lower-temperature sources. Thus, in other galaxies we are  observing the SSSs that are not within or behind regions with high absorption. Furthermore, we preferentially detect the higher temperature sources which also tend to be the brightest SSSs. 
Beyond simply finding more SSSs \citep{RDKongSSS} in external galaxies, we also discovered extensions of the class.

\begin{itemize}

\item{} {\bf QSSs:} External galaxies were found to contain not only SSSs, whose flux falls to zero above $\sim 1$~keV \citep{RDKongQSS}, but also a higher photon-energy extension composed of {\sl quasisoft X-ray sources (QSSs).} QSSs emit photons above  1~keV, but not above $\sim 2$~keV. The estimated values of the effective temerature are typically 
$\sim 100 - 250$~eV.

\item{} {\bf ULSs:} 
When we observe entire galaxies, as {\sl Chandra} and {\sl XMM-Newton} do for many galaxies, we find more bright X-ray sources (XRSs) of every type. This has allowed us to discover rare classes of XRSs. Among the most important of these are ultraluminous XRSs (ULXs).  ULXs are non-nuclear sources with X-ray luminosities above $\sim 10^{39}$~erg~s$^{-1}$.
As is the case with SSSs, the empirical definition of the class means that is likely composed of subclasses, each having a different physical nature \citep[see][for a review]{King23}. Some have been identified as neutron-star systems through the detection of pulsations \citep{IsraelNSULX}. Others are thought to be black-hole accretors \citep{KongRDM101ULS1}. BHs of stellar mass and higher have been considered. Ultraluminous SSSs (ULSs) have been discovered. They constitute a significant fraction (as many as 1 in 7) of all ULXs \citep{Liu11}. 

Some ULXs and ULSs have been observed to either be transient or to pass through phases in which the X-ray spectrum changes \citep{KongRDM101ULS1}. When they are ``on'' they contribute significantly to the soft X-ray flux emitted by the galaxy. It is important to note that ULSs tend to have temperatures at the high end of the SSS temperature range, with values that can be $\sim 100$~eV or higher. Some would be called QSSs.

\item{} {\bf HSSs:} Recently, a low-photon-energy extension of the class of SSSs has been discovered (private communication). These {\sl Hypersoft X-ray sources (HSSs)} exhibit little or no emission above $0.3$~keV. HSSs appear to be sources that have values of $k\, T$ below roughly $20-25$~eV.  Among the ``classical SSSs''--i.e., those found in the Local Group, some have lower limits of $k\, T$ in this range, and one has both a lower and upper limit consistent with HSSs \citep{GreinerSSS}. These classical low-photon-energy sources have typical X-ray luminosities around $10^{37}$~erg~s$^{-1}$. There 
is a likely connection between the classical SSSs and HSSs: when the photosphere of an SSS expand, pushing the emission out of the very nearby regime, it should then be emitting as an HSS. Within the Local Group, two M31 novae have been seen to transition from an SSS phase to an HSS phase as they cool post-explosion. These HSSs have lumiosities in the range $10^{36}-10^{37}$~erg~s$^{-1}.$

Just as with SSSs, however, there appear to be other classes of HSSs that may be ultraluminous. 
The HSSs discoveries outside of the Local group have typical luminosities about ten to about 100 times higher than characteristic of the local SSSs. Such high luminosites cannot be generated by WDs. This suggests that HSSs in other galaxies may be NS or BH accretors. Like the WD accretors whose photoshpere appears to increase in size, the same may be true of NS and BH accretors, where  the source's photosphere and/or the radius of the inner disk increases. 
\end{itemize}

Given these challenges, in this paper we do not assume anything about the nature of SSSs. For example, they may be WDs, NSs, BHs, or stripped stellar cores (WR stars). Instead, we use the term ``SSSs'' to refer to members of the full family of physical model. We compute the ionization due to SSSs having a range of luminosities and temperatures and compute how large the population would have to be in order to contribute significantly to the observed ionization states. 
  
\section{Methods} \label{sec:modeling}

The nebular emission lines depend on two key aspects of the photoionization process: the ionizing radiation field and the characteristics of the ionized gas. This paper focuses on the radiation field powered by SSSs to predict the optical \heiioptic/\hbeta line ratio in star-forming galaxies. We assume that the spectral energy distribution (SED) of SSSs conforms to a blackbody model, which serves as one of the the input for the MAPPINGS photoionization code \citep{Sutherland18}.

We investigate how variations in the input spectra and gas ionization parameters affect the calculated \heiioptic/\hbeta emission line ratio produced by MAPPINGS. The \heii/\hbeta ratio is used instead of the total \heii luminosity because it serves as a robust tracer of the hardness of the ionizing spectrum. The production of \heiioptic requires extremely high-energy photons ($>54.4$ eV), whereas \hbeta originates from lower-energy ionizing photons ($>13.6$ eV). As a result, the \heii/\hbeta ratio directly measures the hardness of the radiation field, making it a crucial diagnostic for distinguishing different ionizing sources. Additionally, \hbeta is a well-calibrated recombination line, and using \heii/\hbeta allows for more robust flux calibration, minimizing observational uncertainties related to instrumental sensitivity, distance, and aperture effects. The small wavelength separation between \heiioptic and \hbeta$\lambda4861$ ensures that dust attenuation affects both lines similarly, effectively canceling out extinction effects and making the ratio less sensitive to dust corrections than absolute HeII luminosity.

The results from these models are compared with observed datasets from both local and high-redshift galaxies to identify the parameter ranges that can reproduce the observations. In this section, we describe the parameters used in the models, discuss the different parameter values tested, and summarize them in Table \ref{tab:variables}. Additionally, we provide a detailed explanation of the SED used in the models and outline the configuration of the parameter grid employed in the MAPPINGS simulations.

\subsection{Ionizing radiation field} \label{ssec:sed}

\subsubsection{BPASS stellar population synthesis}
In star-forming galaxies, stellar populations of main-sequence stars and giants are considered to be the primary source of ionizing photons. The spectral energy distribution of the stellar component changes primarily with stellar age and metallicity \citep[see][for a review]{Gallazzi05}. Several models have been developed to predict the stellar spectra with different prescriptions regarding the initial mass function (IMF), stellar evolution track and spectral library. In these models, the shape of the UV flux beyond the Lyman limit is highly uncertain, depending on the adopted stellar evolution and atmospheric model. These cause systematic uncertainties in the production of $\mathrm{He^{+}}$ ionizing photons. So far, models including only stellar components have not been able to produce sufficient UV flux to reproduce the observed \heii flux in star-forming galaxies \citep{Xiao18}. 

We adopt the stellar spectral energy distribution (SED) from BPASS 2.2, which incorporates binary stellar evolution. The inclusion of binaries is important because it produces harder ionizing spectra than single-star populations, improving agreement with observed UV and optical emission lines \citep{Xiao18}. However, while supersoft X-ray sources (SSSs) arise from binary evolution, their spectra are not included in BPASS. BPASS 2.2 models the evolution of single and binary stars using the \textsc{STARS} stellar evolution tracks \citep{Eggleton71, Eldridge08}. The synthetic stellar spectral library is primarily based on BaSel v3.1 \citep{Westera02}, complemented by \citet{Lejeune98} for lower-metallicity stars. Wolf-Rayet (WR) stars are modeled using the Potsdam PoWR theoretical library \citep{Hamann06, Sander15}, while O stars are computed with the WM-Basic code \citep{Pauldrach01}, which adopts the stellar atmosphere grid from \citet{Smith02}.

\begin{figure}
    \centering
    \includegraphics[width=1.0\linewidth]{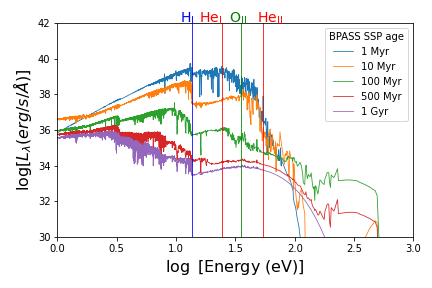}
    \caption{The spectral energy distribution of binary stellar populations from BPASS with ages of 1Myr - 1Gyr. The vertical lines show the ionization potential of $\mathrm{H_{I}\ (13.59 eV),\ He_{I}\ (24.59 eV),\ O_{II}\ (35.12 eV)}$ and \heii $\mathrm{(54.42 eV)}$}
    \label{fig:bpass}
\end{figure}

BPASS provides nine IMF options. We use the one that resembles the \citet{2001Kroupa} IMF which consists of a double power law: a slope of $\alpha_1 = -1.30$ for stellar mass range of $0.1$ to $0.5 \Msun$, and a steeper slope of $\alpha_2 = -2.35$ for a higher stellar mass range from $0.5 \Msun$ to $100 \Msun$. We assumed a stellar metallicity of $Z = 0.40\ \Zsun$. Figure \ref{fig:bpass} shows the SED of stellar populations from BPASS with our chosen metallicity and different stellar population ages, ranging from 1 Myr to 1 Gyr, normalised for a total stellar mass of $10^6$ \Msun. Vertical lines in the figure marks ionization potential of different species: $\mathrm{H_{I},\ He_{I},\ O_{II}}$ and \heii.

A key objective of this study is to compare the HeII ionization contributed by stellar populations to that produced by SSSs. Constructing a galaxy's SED from BPASS requires an assumption about its star formation history (SFH). While real galaxies experience star formation in bursts over extended timescales, nebular emission lines are primarily shaped by the extreme UV radiation from short-lived massive stars. This allows a simplified approach to approximate the dominant ionizing photon production. To maximize the predicted \heii emission from stars, we adopt an instantaneous burst SFH, where $10^6\ \Msun$ of stars form in a single epoch. Figure \ref{fig:bpass} shows that stellar populations at an age of 1 Myr produce the highest flux at \heii ionization potential. As stellar populations age, the \heii-ionizing flux declines rapidly because the most massive stars, which are the primary sources of extreme UV radiation, leave the main sequence. For this reason, we select 1 Myr as our representative stellar age, ensuring we model the peak \heii-ionizing output from massive stars and directly test whether such populations alone can account for the observed HeII flux.

In Figure \ref{fig:bpass}, we see that the youngest stellar population with an age of 1 Myr produces the highest flux in the energy range corresponding to the ionization potential of the four species shown. The flux of ionizing photons decreases as the stellar population age increases because the most massive stars which produce these photons have died. Because of this, we chose the 1 Myr SED normalised for a $10^6 \Msun$ population as the basis for our photoionization model. We note that old stellar populations (100-500 Myr) produce higher flux in the energy range above $100$ eV compared to the younger populations, which account for the influence of binaries in the later stage of stellar population. For example, white dwarfs (WDs) only start to form after about 100 Myr and the formation rate increases after that. SSSs and other X-ray binaries are part of these population as well but their X-ray emission do not appear here. 

At later times (100–500 Myr), the SED of stellar populations exhibits a blue tail extending beyond 100 eV, though this is largely due to binary interactions that extend the production of ionizing radiation \citep{Wofford16, D'agostino19}. In binary systems, stars can gain mass through Roche-lobe overflow, altering their evolutionary pathways and allowing them to behave as if they were more massive and younger than their true age \citep{Dray07}. Similarly, stellar mergers create rejuvenated stars that sustain high-energy emission over longer timescales \citep{Schneider16}. Binary interactions can also strip red supergiants of their hydrogen-rich envelopes, exposing hot helium cores and leading to the formation of Wolf-Rayet (WR) or helium stars. Unlike in single-star populations, where WR stars primarily arise due to strong stellar winds, binary evolution allows them to persist at much later ages, extending the ionizing lifetime of the population \citep{Xiao18}.

Despite these factors, even BPASS models that account for binary evolution fail to produce enough ionizing photons to explain observed \heii emission in star-forming galaxies. A crucial limitation of BPASS v2.2 is that it does not explicitly include accreting binaries, such as X-ray binaries and SSSs. Because of this, BPASS alone fails to reproduce the observed intensity of \heii \citep{Xiao18, Stanway19}. Many studies attempt to correct for this by introducing X-ray binary populations in an \textit{ad-hoc} manner \citep{Schaerer19, Senchyna20, Garofali24}, but such modifications highlight the need for additional ionizing sources, such as SSSs, to fully explain \heii production in galaxies.

\subsubsection{Blackbody to represent SSSs}

The X-ray spectra of SSSss are commonly modelled as blackbody spectra. To approximate the ionizing radiation field emitted by SSSs, we input the blackbody spectrum with seven different temperatures, ranging from 5eV ($T \approx50$kK) to 100eV ($T \approx 10^6$K). Our primary focus is on testing the reproduction of highly ionized lines using a blackbody representing SSSs, and we do not aim to consistently match the low ionization lines, which have previously been successfully predicted using SED from stellar population synthesis. In any case, SSSs are not expected to contribute significantly in this range.

\begin{figure}
    \centering
    \includegraphics[width=1.0\linewidth]{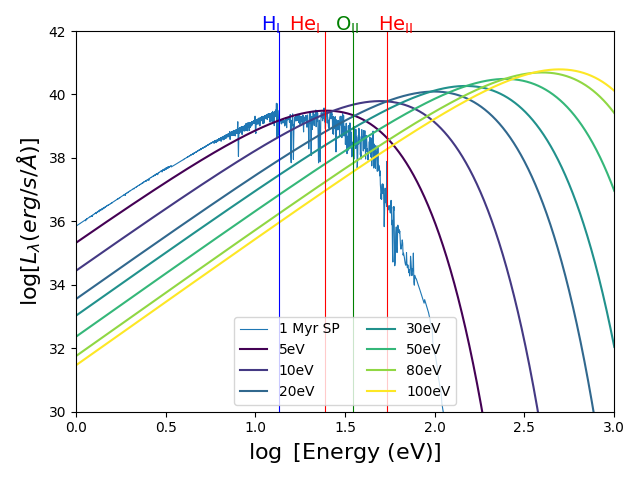}
    \caption{The spectral energy distribution of a 1 Myr stellar population from BPASS (blue, showing emission and absorption lines) and blackbody spectra of $kT=5, 10, 20, 30, 50, 80, 100$ eV (purple to yellow, as indicated in the legend). The total luminosity of each blackbody is set equal to the luminosity of the 1 Myr stellar population from BPASS, assuming a total stellar mass of $10^6$ \Msun. The vertical lines show the ionization potential of $\mathrm{H_{I}\ (13.59 eV),\ He_{I}\ (24.59 eV),\ O_{II}\ (35.12 eV)}$ and \heii $\mathrm{(54.42 eV)}$}
    \label{fig:bpass_bbody}
\end{figure}

Figure \ref{fig:bpass_bbody} shows blackbody spectra with temperature $kT = 5, 10, 20, 30, 50, 80$ and 100 eV compared to the spectra of 1 Myr stellar population from BPASS with a total stellar mass of $10^6$ \Msun. The color varies from purple to yellow as the temperature varies from 5 to 100 eV. In this figure, we assume that the bolometric luminosity of each blackbody equals to that of the stellar population. Again, the vertical lines mark the ionization potential of $\mathrm{H_{I},\ He_{I},\ O_{II}}$ and \heii. The figure shows that the inclusion of a blackbody source can be important in modelling high ionization species. The contribution of a $5 - 10$ eV blackbody is roughly equal to the stellar component for $\mathrm{He_{I}}$ ionization. While young stellar spectra produce more ionizing spectra for low ionization species like $\mathrm{H_{I}}$, a blackbody is gaining more importance for species with higher ionization energy like $\mathrm{O_{II}}$ and \heii. 

We use a blackbody with varying temperatures as input to our photoionization model to compute the resulting \heii and \hbeta line fluxes. In each model run, we fix the bolometric luminosity at $L = 10^{39} \mathrm{erg/s}$, consistent with the average luminosity of a single SSS. The ionization parameter of the nebular gas is included as an additional variable to explore its impact on \heii production. We present the various parameters in Table \ref{tab:variables} and discuss the effect of each parameters in Section \ref{ssec:blackbody}. 

\subsubsection{Combination of stellar spectra and blackbody}
We combine the stellar spectra from BPASS with blackbody spectra to represent the contribution of SSSs to the ionizing radiation field in galaxies. The 1 Myr stellar population from BPASS is used to approximate the stellar contribution to ionization, as young, massive stars are the dominant producers of ionizing photons in galaxies. As shown in Figure \ref{fig:bpass}, the ionizing photon flux peaks at early stellar ages, making the 1 Myr population a reasonable proxy for the primary sources of \heii ionization in typical star-forming environments. To explore the impact of SSSs, we incorporate blackbody spectra at seven different temperatures, as presented in Table \ref{tab:variables} and illustrated in Figure \ref{fig:bpass_bbody}. This range of temperatures allows us to assess the spectral diversity of SSSs and their effect on the ionizing radiation field.

In the combined spectra, we examine the contribution of the blackbody component to the BPASS stellar spectra by varying its bolometric luminosity fraction relative to the stellar population. The applied luminosity fractions are 0 (pure stellar), 0.01, 0.1, 1, and 10, as listed in Table \ref{tab:variables}. The total system luminosity is fixed at $L = 10^{39} \mathrm{erg/s}$ to maintain a consistent energy scale across all models.

\subsection{Photoionization model grid} \label{ssec:mappings}

We use the input ionizing spectra, described in Section \ref{ssec:sed}, for the \textsc{MAPPINGS V}\footnote{https://mappings.anu.edu.au/code/} code \citep{Nicholls14, Sutherland18} to generate a grid of photoionisation models. \textsc{MAPPINGS} includes heating, ionization and shocks within a gas nebula to compute the resulting nebular spectra (continuum and emission line flux). It takes into account the input ionizing spectrum, gas properties of the nebula and a dust treatment \citep{Groves04}. The atomic data used in the latest version comes from the CHIANTI version 8 \citep{DelZanna15}.

%As mentioned in Section \ref{ssec:sed}, we explore two kinds of ionizing input spectra: blackbody radiation to represent the SSS, and a combination of blackbody and BPASS stellar population synthesis to illustrate the overall spectra of galaxies with SSS. In model variants with blackbody radiation, we vary the temperatures to be $kT = 5, 10, 20, 30, 50, 80, 100 \mathrm{eV}$. This temperature range covers the observed temperature of various soft X-ray sources. The total luminosity of the blackbody is set to be in the range of $10^{36}$ to $10^{39}$ $\mathrm{erg s^{-1}}$ to match the observed luminosities of SSS. 

%The second model variant of ionzing spectra use a combination of a blackbody and a BPASS spectra. We use a similar temperature range with the pure blackbody model but fix the total luminosity of the combined system to be $10^{39}$ $\mathrm{erg s^{-1}}$. In this model variant, we explore the effect of changing blackbody contribution to the combined spectra. We applied five blackbody scaling with a fix BPASS luminosity: 0.1, 1, 10, 100 and 1000. 

\begin{table*}[]
    \centering
    \begin{tabular}{c c c c}
        \hline
        Ionizing spectra & Parameters & Values & Unit \\
        \hline
        Blackbody & $kT$ & 5, 10, 20, 30, 50, 80, 100 & eV \\
         %& Bolometric luminosity & $10^{36}$, $10^{37}$, $10^{38}$, $10^{39}$ & $\mathrm{erg\ s^{-1}}$ \\
         & Ionization parameters ($\log U$) & -4, -3.5, -3, -2.5, -2, -1.5, -1 & - \\
         \hline
        Blackbody $+$ BPASS  & $kT$ & 5, 10, 20, 30, 50, 80, 100 & eV \\
         %& Bolometric luminosity & $10^{39}$ & $\mathrm{erg\ s^{-1}}$ \\
         & Blackbody luminosity fraction & 0, 0.01, 0.1, 1, 10 & -\\
         & Ionization parameters ($\log U$) & -4, -3.5, -3, -2.5, -2, -1.5, -1 & - \\
        \hline

    \end{tabular}
    \caption{Varying parameters of the model grid}
    \label{tab:variables}
\end{table*}

When running the photoionization model, we set the gas-phase metallicity to be $12 + \log (\mathrm{O/H}) = 8.47$. We use the solar abundance from \citet{Anders89} and the depletion factors from \citet{Jenkins14} for each element, with a logarithmic base depletion of 1.50 for iron. These values are chosen to mimic the conditions in the diffuse interstellar medium of the Milky Way. MAPPINGS incorporates a dust treatment with grain size distribution following \citet{Mathis77}, without allowing for grain destruction. The depletion fraction of polycyclic aromatic hydrocarbons (PAHs) is set at $0.30$.

Ionization parameters ($Q, \mathrm{cm}\ \mathrm{s^{-1}}$) are adopted with values ranging from $\log \mathrm{Q} = 6.5$ to $9.5$ in 0.5 dex increments, corresponding to dimensionless ionisation parameters between $\log U = -4$ and $-1$. The ranges are selected to align with observed values for local starburst galaxies \citep{Kewley01, Rigby04}, as well as the elevated ionization parameters found in high-redshift galaxies.

We assume a closed-spherical geometry, where radiation is confined within the nebula, and all ionizing photons interact with the gas. We use the radiation-bounded assumption, meaning the gas absorbs all ionizing photons. The gas pressure is held constant at $P/k = 10^5\ \mathrm{cm^{-3}\ K}$, with an initial temperature of $T = 10.000\ \mathrm{K}$, corresponding to initial density of $10\ \mathrm{cm^{-3}}$. The chosen value is comparable to the observed density of $26.8^{+0.2}_{-0.2}\ \mathrm{cm^{-3}}$ for local galaxies in the COSMOS-[$\mathrm{O_{II}}$] sample \citep{Kaasinen17}. 

\subsection{Observational sample} \label{ssec:observation}
We compare our model predictions with the observed \heii/\hbeta line ratio detected in galaxies at both low and high redshifts. 
\subsubsection{Low-redshift sample (SDSS)}

Our low-redshift data consist of strong emission-line galaxies from SDSS DR7 \citep{Abazajian09}, specifically those with significant \heii\ emission as compiled by \citet{Shirazi12}. This sample is based on the parent galaxy catalog from \citet{Brinchmann04}, who demonstrate that an $\mathrm{S/N}>3$ threshold for key emission lines (including $\mathrm{H_\beta}$, $\mathrm{[O_{III}]}$, $\mathrm{[N_{II}]}$, and $\mathrm{H_\alpha}$) effectively reduces the number of galaxies with unreliable (negative) flux measurements, which could bias the resulting BPT classification. Subsequently, \citet{Shirazi12} imposed a more stringent criterion ($\mathrm{S/N}>5$) specifically on the detection of \heii, further ensuring reliability. The resulting robustly classified star-forming subsample consists of 189 galaxies at redshifts $z<0.2$. This selection is unlikely to significantly bias our distribution, as the conservative $\mathrm{S/N}>5$ threshold specifically for \heii\ substantially mitigates contamination from positive outliers. The blue regions in Figures~\ref{fig:He-Qvar} and \ref{fig:He-combo_L} illustrate the $1\sigma$ distribution of the observed \heii/\hbeta\ ratio derived from this low-redshift sample.

\subsubsection{High-redshift sample (JWST/NIRSpec)}
Our high-redshift observational data are drawn from JWST/NIRSpec spectra publicly available through the Dawn JWST Archive (DJA). We begin by selecting galaxies from the initial DJA spectroscopic catalog release\footnote{\url{https://dawn-cph.github.io/dja/blog/2023/07/18/nirspec-data-products/}} (v0) with significant detections ($\mathrm{S/N}>3$) in both \heii\ and $\mathrm{H_\beta}$ lines and robust quality flag (grade = 3), following criteria established by \citet{Heintz23}. Applying a stricter $\mathrm{S/N}>5$ threshold for both lines substantially reduces our sample size to only three galaxies, limiting our ability to conduct meaningful statistical analysis. Therefore, to preserve statistical significance, we adopt the less restrictive but still widely used criterion of $\mathrm{S/N}>3$ for both emission lines.

To quantify potential bias introduced by this choice, we compare the average \heii/\hbeta\ ratios for the two thresholds. The stricter $\mathrm{S/N}>5$ criterion yields an average log(\heii/\hbeta) of approximately $-0.92$, while the adopted $\mathrm{S/N}>3$ criterion gives an average of about $-0.75$. Thus, relaxing the selection criterion introduces an upward bias of approximately $18\%$. Although this positive bias slightly elevates the average emission line ratio by selecting stronger emission-line detections, it does not alter the overall interpretation of our analysis, which focuses primarily on the range and distribution rather than solely on the average line ratios.

We then retrieved the spectra from the v2 data release\footnote{https://dawn-cph.github.io/dja/blog/2024/03/01/nirspec-extractions-v2/} and refitted them using the \textsc{msaexp} pipeline \citep{msaexp}, obtaining new line measurements. Applying our adopted $\mathrm{S/N}>3$ criterion yields eight galaxies spanning redshifts $z\approx2-6$. These observations comprise the orange region in Figures \ref{fig:He-Qvar} and \ref{fig:He-combo_L}. Our measurements for these eight measurements are presented in Table \ref{tab:nirspec}.

\begin{table*}[]
    \centering
    \begin{tabular}{c c c c c c}
        \hline
        program & disperser-filter & ID & redshift & \heii & \hbeta \\
         & & & & line flux & line flux \\
         & & & & ($10^{-20} \mathrm{erg/s/cm^2}$) & ($10^{-20} \mathrm{erg/s/cm^2}$)  \\
        \hline
        ceers & g140m-f100lp & 1345\_3506 & 2.0552 & $139.73\pm24.01$ & $2841.81\pm75.20$\\
        ceers & g395m-f290lp & 1345\_355 & 6.0996 & $55.83\pm14.37$ & $134.46\pm16.69$ \\
        gds-deep & g395m-f290lp & 1210\_13176 & 5.9352 & $12.29\pm2.51$ & $144.09\pm4.10$ \\
        gds-deep & g395m-f290lp & 1210\_13620 & 5.9168 & $18.29\pm5.04$ & $49.69\pm4.84$ \\
        gds-deep & prism-clear & 1210\_13176 & 5.944 & $15.95\pm2.17$ & $218.59\pm4.23$ \\
        gds-deep & prism-clear & 1210\_9414 & 5.8928 & $30.22\pm4.46$ & $62.26\pm4.91$ \\
        jades-gds-wide & prism-clear & 1180\_1594 & 4.2776 & $71.02\pm22.89$ & $75.49\pm23.41$ \\
        snh0pe & g235m-f170lp & 4446\_19 & 3.951 & $33.79\pm10.58$ & $716.06\pm18.13$ \\
         & 
    \end{tabular}
    \caption{Line measurement of galaxies with reliable \heii nebular emission from NIRSpec v2 data release}
    \label{tab:nirspec}
\end{table*}

%three nearby extremely metal poor galaxies (EMPGs) with \heii detection, as documented by \citet{Umeda22}. Specifically, these galaxies are J1631+4426, J104457, and I Zw 18 NW. Local EMPGs are considered as counterparts to young galaxies at high-redshift due to their remarkably low metallicity (below $10 \%$ solar values), low mass and high specific star formation rate (sSFR). The scarcity of WR stars in these galaxies makes it unlikely that they are the primary source of the detected \heii. Consequently, we explore the possibility of the SSS as the origin of \heii. In the BPT diagram, these galaxies lies far below the \citet{Kauffmann03} line, indicating their ionization originates from stellar sources.

%We complement the observation of nearby samples with emission lines detected in high-redshift galaxies using NIRSPEC instrument in JWST

%We complement the observation of nearby samples with three galaxies at $z > 7.5$: S04590, S06355, and S10612 from \citet{Katz23}. These galaxies do not have \heii detection, but they have strong [OIII]$\lambda4363$ flux, which is a high-ionization line. -- do we want to add 4363 line ratio as well

\section{Results} \label{sec:results}

\subsection{\heii emission from supersoft X-ray source}
\label{ssec:blackbody}

We model the spectra as a blackbody to test whether SSSs can account for the elevated \heiioptic emission lines observed in galaxies at both low and high redshifts. We use the \heiioptic/\hbeta instead of the absolute \heiioptic luminosity to minimize biases related to intrinsic brightness, stellar mass, distance, and observational uncertainties, as these factors affect both lines similarly. Additionally, the proximity of \heiioptic and \hbeta$\lambda4861$ in wavelength space minimizes the impact of dust attenuation, ensuring that variations in the \heii/\hbeta ratio primarily reflect differences in ionizing photon production, with high-energy photons (54.4eV) producing \heii and lower-energy photons (13.6eV) responsible for \hbeta. The \heii/\hbeta ratio may also be less affected by intergalactic medium (IGM) absorption compared to the absolute HeII flux, since both lines are close enough in wavelength that their optical depths are similarly impacted by intervening neutral gas.

The main parameters varied to study their effect on the \heii/\hbeta ratio are the blackbody temperature and ionization parameter, as discussed in Section \ref{ssec:sed} and Table \ref{tab:variables}. Additionally, we explored the effect of bolometric luminosity and found that it does not significantly impact the \heii/\hbeta ratio. This result is expected because increasing the bolometric luminosity proportionally increases the total line flux for both \heii and \hbeta, meaning their ratio remains largely unchanged.

\subsubsection{Effect of varying blackbody temperatures on \heii emission} \label{sssec:temperature}

The first step in our analysis is to assess whether stellar populations alone can account for the observed \heii/\hbeta ratios in galaxies. The crosses in Figure \ref{fig:He-Qvar} represent the \heii/\hbeta line ratio predicted by BPASS-only models, using the most ionizing stellar population (1 Myr, no blackbody). The color of each cross indicates the ionization parameter.

The BPASS-only model fails to reproduce the observed \heii/\hbeta ratios, falling 1 dex below the SDSS sample and 2 dex below the NIRSpec sample. This result confirms that even the most extreme stellar populations in BPASS do not produce enough hard ionizing photons to account for the observed \heii emission, demonstrating the need for an additional ionizing source

To explore alternative sources, we introduce blackbody radiation to represent the contribution of SSSs. The blackbody temperature plays an important role in shaping the ionizing spectrum, as illustrated in Figure \ref{fig:bpass_bbody}. A blackbody spectrum with $kT=10 \mathrm{eV}$ peaks around $\log(E/\mathrm{eV}) = 2$, whereas a $kT=100 \mathrm{eV}$ spectrum peaks at $\log(E/\mathrm{eV}) = 3$. This confirms that higher kT shifts the peak energy to harder photons

Figure \ref{fig:He-Qvar} shows how varying kT impacts the \heii/\hbeta ratio. The blue shaded region represents the 16th to 84th percentile range of the \heii/\hbeta line ratio from the local SDSS sample, while the orange shaded region represents the same percentile range for the high-redshift NIRSpec sample. The variation in the ionization parameter is represented by the color bar to the right of the figure. 

Increasing the temperature from $kT = 5 \mathrm{eV}$ to $20 \mathrm{eV}$ results in a 2-dex increase in the \heii/\hbeta ratio. However, further increasing the temperature to $100$ eV leads to a 0.5-dex decrease, indicating that HeII production efficiency peaks around 20–40 eV. This peak aligns well with the observed temperature range of CAL83, the only known SSS with detected \heii emission in its surrounding nebula \citep{RemillardNeb, WoodsGilNeb}. CAL83 has a temperature range of 20-50 eV, further supporting this model. In contrast to BPASS-only models, pure blackbody models fully cover the observed \heii/\hbeta ranges in both SDSS and NIRSpec samples, with temperatures ranging from $kT = 5-100$ eV.

These results show that even the  most ionizing young stellar populations in BPASS fail to reproduce observed HeII emission, reinforcing the need for an additional ionizing source in the soft X-ray regime (kT $< 100$ eV). SSSs provide a physically motivated candidate for this missing ionizing population, making them a compelling explanation for HeII production in both local and high-redshift galaxies.

\subsubsection{Effect of ionization parameters in the \heii emission} \label{sssec:ionizationparameter}
\begin{figure}
    \centering
    \includegraphics[width=1.0\linewidth]{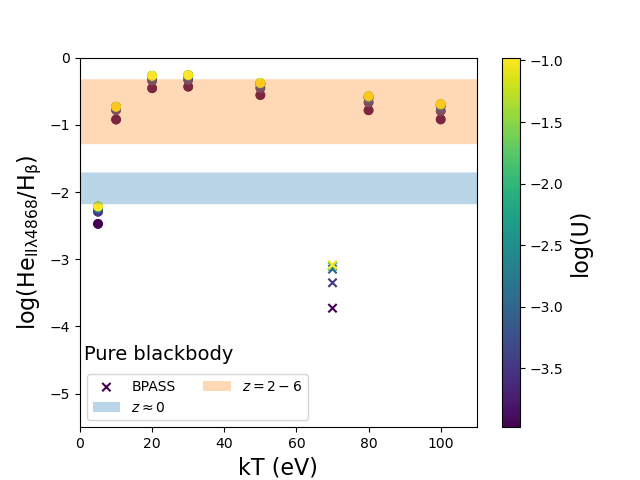}
    \caption{\heii/\hbeta as a function of blackbody temperature for the MAPPINGS model with pure blackbody spectra as an input. The bolometric luminosity is kept constant at $L = 10^{39} \mathrm{erg/s}$. Symbols with different colors marks different ionization parameters as indicated in the colorbar. Crosses marks the MAPPINGS output from the 1 Myr stellar population synthesis from BPASS with a metallicity of $0.4\ \Zsun$. The gas-phase metallicity is assumed to be similar to the stellar metallicity and the ionization parameter of the BPASS model is kept at a similar range with the pure blackbody model. The blue shaded region shows the 16th and 84th percentile of the \heii/$\mathrm{H_\beta}$ ratio observed in the SDSS sample and the orange shaded regions shows the 16th and 84th percentile in the NIRSpec sample.}
    \label{fig:He-Qvar}
\end{figure}

One of the important properties of nebular gas that affects emission lines is the ionization parameter. Ratios of strong emission lines, like [OIII]$\lambda\lambda4959,5007$ and [OII]$\lambda\lambda3727,3729$, are known to be sensitive to the ionization parameter and can be used as diagnostics to trace it. In Figure \ref{fig:He-Qvar}, symbols are colored based on the ionization parameter $\log (U)$, with lighter colors indicating higher ionization parameters. The crosses represent the \heii/\hbeta line ratio from BPASS models across the same range of ionization parameters.

In Figure \ref{fig:He-Qvar}, the blackbody model shows 0.4-dex increase in the \heii/\hbeta line ratio with an increasing ionization parameter from $\log(U) = -4$ to $-2.5$. The increase is more pronounced for the BPASS model, reaching approximately $0.8$ dex over the same range. However, the rise in the \heii/\hbeta line ratio is not linear with increasing ionization parameter; it slows significantly at higher values of $\log(U)$. When the ionization parameter becomes sufficiently high, nearly all atoms in the nebular gas are already highly ionized, causing the \heii/\hbeta line ratio to plateau. 

Figure \ref{fig:He-Qvar} shows that even though the \heii/\hbeta line ratio from the pure BPASS model is sensitive to changes in the ionization parameter, it does not reach the observed level even after the plateau is reached. In contrast, the blackbody model succesfully provides the ionization needed for the production of \heii in a typical ISM.

\subsection{\heii emission from star-forming galaxies with supersoft X-ray sources}
\label{ssec:combination}
A blackbody with a total luminosity of $L = 10^{39} \mathrm{erg/s}$ can reproduce the observed \heii/\hbeta line ratio in galaxies. However, to fully explain the detectability of \heii lines in star-forming galaxies, it is important to assess the relative contributions of the blackbody component to the stellar populations. In this section, we examine combined spectra from a 1 Myr BPASS stellar population and a blackbody spectrum. The bolometric luminosity fraction of the blackbody relative to the stellar population is defined as $f_{bb} = L_{bol,bb} / L_{bol,\mathrm{BPASS}}$. We vary $f_{bb}$ across $0, 0.01, 0.1, 1$ to $10$, while keeping the total bolometric luminosity of the stellar population fixed at  $L = 10^{39} \mathrm{erg/s}$ and the ionization parameter constant at $\log U = -3$. 

Figure \ref{fig:He-combo_L} shows the \heii/\hbeta line ratio as a function of blackbody temperature for models using the combined spectra. Different colors and symbols represent models with varying blackbody contributions, as indicated in the legend. The red cross marks the \heii/\hbeta line ratio produced by BPASS-only model ($f_{bb}=0$). 

For $kT = 5$ eV, the combination of BPASS and blackbody spectra produces results similar to the pure stellar population model when $f_{bb} < 0.1$ (represented by the leftmost blue and green markers in Figure \ref{fig:He-combo_L}). This suggests that at low blackbody temperatures, the blackbody contribution is negligible, and \heii ionization is primarily powered by stellar populations.

As $f_{bb}$ increase, the contribution of the blackbody becomes more significant. For example, at $f_{bb} > 0.1$ at $kT=5\mathrm{eV}$, the resulting \heii/\hbeta line ratio is nearly 1 dex higher than that produced by the the pure BPASS model, highlighting the importance of the blackbody component in elevating the \heii/\hbeta ratio. However, the \heii/\hbeta ratios produced by these models are slightly below the ratio observed in SDSS galaxies (blue shaded region), suggesting that higher blackbody temperatures are required to fully reproduce the observed ratios.

For blackbody temperatures in the range $kT = 10-50$ eV, even the lowest blackbody luminosity fraction ($f_{bb} = 0.01$, represented by blue circles in Figure \ref{fig:He-combo_L}) matches the \heii/\hbeta ratio observed in SDSS galaxies. Similarly, a model with $f_{bb} = 0.1$ (green triangles) reproduces the SDSS-observed \heii/\hbeta ratio when using a blackbody with $kT = 5-10$ eV.

A sharp increase in \heii/\hbeta ratio is seen as the blackbody temperature rises from $5$ eV to $20$ eV, after which the ratio declines slightly. In the $kT = 10-50 \mathrm{eV}$ temperature range, a model with $f_{bb} = 1$ (blackbody luminosity equal to the stellar luminosity, represented by purple diamonds in Figure \ref{fig:He-combo_L}) matches the \heii/\hbeta ratios observed in high-redshift galaxies (NIRSpec, orange shaded region). Increasing the blackbody contribution to $f_{bb} = 10$ (grey squares) achieves similar results.

The efficiency of a blackbody in elevating the \heii/\hbeta ratio decreases in models with a $kT = 80-100$ eV. In this temperature range, the \heii/\hbeta ratios produced are $0.2$ to $0.5$ dex lower than those from models with a $50$ eV blackbody. At these higher temperatures, the model with the lowest luminosity fraction ($f_{bb} = 0.01$) fails to reproduce the SDSS observations. Instead, the SDSS galaxies are better matched by models with $f_{bb} = 0.1$ (green triangles). However, models with higher blackbody contribution ($f_{bb} = 1-10$) continue to reproduce the \heii/\hbeta ratios observed in high redshift galaxies (NIRSpec, orange region).

\section{Implications of SSS Contributions to \heii Ionization}
\label{sec:discussion}

\subsection{The need for SSSs}
Our results demonstrate that stellar populations alone, even under the most favorable conditions (e.g., 1 Myr BPASS models), consistently underpredict the observed \heii/\hbeta ratios in both local and high-redshift galaxies. As shown in Figure~\ref{fig:He-Qvar}, the BPASS-only models fall $\sim1$ dex below the \heii/\hbeta values observed in the SDSS sample and $\sim2$ dex below those seen in JWST/NIRSpec galaxies. This discrepancy highlights the need for an additional source of hard ionizing radiation, particularly one capable of producing photons with energies above 54.4 eV, which are required to ionize He$^+$.

To explore this, we introduced a blackbody component representing the contribution of SSSs and combined it with BPASS stellar spectra. By varying both the blackbody temperature ($kT$) and its relative bolometric luminosity contribution ($f_{bb}$),  we identified the parameter space that successfully reproduces the observed \heii/\hbeta ratios. For the SDSS sample, which has a median $\log$(\heii/\hbeta) of approximately $-2$, we find that the required SSS contribution is modest, and can be made up from several parts of the parameter space:
\begin{itemize}
    \item $kT = 10$–$30$ eV with $f_{bb} = 0.01$ ($0.9\%$ of the total ionizing luminosity), 
    \item a lower temperature of $kT = 5$–$10$ eV with $f_{bb} = 0.1$ ($9\%$ of the total ionizing luminosity),
    \item a high temperature of $kT = 100$ eV with similar contribution $f_{bb} = 0.1$.
\end{itemize}
In contrast, to reproduce the elevated \heii/\hbeta ratios observed in NIRSpec galaxies ($\log$(\heii/\hbeta) $ \approx -1.3$ to $-0.3$), a much higher SSS contribution is required. Models match these observations only when the blackbody temperature ranges from $kT = 10$–$100$ eV and the blackbody contribution reaches $f_{bb} = 0.1$–$10$ (i.e., $9–90\%$ of the total ionizing luminosity). These estimates are summarized in Figure~\ref{fig:He-combo_L}, which shows that while modest SSS contribution can account for local \heii emission, a substantially larger SSS population—or equivalently, a greater integrated contribution from soft X-ray sources—is needed to explain the stronger HeII emission observed in early galaxies.

The required increase in soft ionizing contribution from local to high-redshift galaxies is also physically motivated. Although our model does not assume a specific origin for the SSSs, their numbers are expected to scale with the star formation activity of the host galaxy \citep{Galiullin21}. Observations of nearby galaxies (see Table~\ref{tab:sss_counts}) show that SSS detections are significantly more common in star-forming spirals than in passive ellipticals, suggesting a link between recent star formation and SSS production. This is consistent with theoretical expectations that SSSs arise from post-main-sequence binary evolution, which is enhanced in younger stellar populations. Given that galaxies at $z \sim 2$–6 typically exhibit elevated star formation rates and bursty star formation histories, the presence of larger SSS populations in high-redshift systems is plausible. This trend provides an intuitive explanation for why the required SSS contribution ($f_{bb}$) increases at high redshift, and motivates a more quantitative analysis of SSS population sizes in the following section.

This analysis provides a phenomenological constraint on the required ionizing component. We do not assume a specific physical nature for the SSSs, but simply require that they emit soft X-ray photons in the 5–100 eV range, and contribute a specified fraction of the ionizing luminosity. In this way, the modeling framework is agnostic to the exact origin of the SSSs, whether they be accreting white dwarfs, X-ray binaries, stripped stars, or other compact sources. The key requirement is that they efficiently produce photons at or above the He$^+$ ionization threshold, which standard stellar population models fail to supply.

\begin{table*}[]
    \centering
    \begin{tabular}{c|c|c|c|c|c}
    \hline
        Morphology & Source & $M_* (10^{10} \Msun)$ & SFR (\Msun $\mathrm{yr^{-1}}$) & SSS count \citep{Liu11} & SSS count \citep{Wang16} \\
        \hline
        \hline
        Spiral & M51 & 4.5 & 2.9 & 27 & 28 \\
                & M81 & 5.4 & 0.5 & 11 & 13 \\
                & M83 & 3.0 & 3.1 & 28 & 42 \\
                & M101 & 2.3 & 2.9 & 36 & 35 \\
        \hline
        Lenticular & Cen A & 4.9 & 0.8 & 8 & 11\\
                   & NGC 3115 & 6.9 & - & 0 & 1 \\
        \hline
        Elliptical & NGC 3379 & 5.8 & - & 6 & 4 \\
                   & NGC 4278 & 6.0 & - & 2 & 3 \\
                   & NGC 4697 & 6.0 & - & 10 & 8 \\
    \hline
    \end{tabular}
    \caption{Properties of galaxies with SSS detection. Morphology, stellar mass and SFR are adopted from \citet{Galiullin21}. SSS counts are derived using Chandra catalogs of \citet{Liu11} and \citet{Wang16}.}
    \label{tab:sss_counts}
\end{table*}

\subsection{Estimating the Required SSS Population}
\label{ssec:estimate}
Having identified the blackbody parameters required to match observed He\,\textsc{ii}/H$\beta$ ratios, we now estimate the number of SSSs needed to contribute the required ionizing luminosity in galaxies. While our model does not assume a specific physical identity for the SSSs, this calculation provides a useful benchmark for assessing whether the inferred contribution is observationally plausible.

In our combined models, we adopted a reference total ionizing luminosity of $L_{\mathrm{ion}} = 10^{39}$ erg s$^{-1}$ to estimate the fractional contribution from SSSs. However, this value can vary significantly depending on the star formation history of the galaxy. We constructed a toy galaxy model using a delayed-$\tau$ star formation history ($\tau = 10$ Gyr), representative of a Milky Way–like system. This star formation history yields a total stellar mass of $\sim 10^{10} \Msun$ and an integrated bolometric luminosity of $L_\mathrm{bol} \sim 3 \times 10^{43}$ erg s$^{-1}$. In the model, the stellar population at 1 Myr contributes approximately 10\% of the total bolometric luminosity, or $L_\mathrm{1Myr} \sim 3 \times 10^{42}$ erg s$^{-1}$. More importantly, this young population alone recovers nearly all of the galaxy's ionizing photon output, justifying its use as a benchmark for modeling sources of He$^+$-ionizing photons.

To estimate the number of SSSs needed to supply a fraction $f_{bb}$ of the ionizing luminosity, we adopt a representative luminosity of $L_{\mathrm{SSS}} = 10^{38}$ erg s$^{-1}$. The number of SSSs per galaxy is then given by:
\begin{equation}
    N_\mathrm{SSS} = f_{bb} \frac{L_{1\mathrm{Myr}}}{L_\mathrm{SSS}}.
\end{equation}

For $f_{bb} = 0.01$--0.1, as required to reproduce He\,\textsc{ii} ionization in local SDSS galaxies, our model yields a population of approximately 3,000 SSSs per galaxy. This serves as a useful fiducial estimate and is consistent with intrinsic SSS population sizes inferred from observations of nearby star-forming galaxies.

Chandra observations detect $\sim$10--40 SSSs per nearby star-forming galaxy (Table~\ref{tab:sss_counts}). These counts, however, are strongly affected by absorption, as most SSSs emit primarily below 100 eV where the photoelectric cross-section of neutral gas is very large. The probability of absorption decreases steeply with photon energy \citep{Morrison83, Wilms00}, so hard X-rays are much less affected. Depending on the source temperature, luminosity, and ISM column density, detection efficiencies of SSSs in star-forming galaxies are likely in the range of 0.1\%--1\% \citep{DiStefano94}. For a typical observed number count, this implies an intrinsic population of $\sim$3,000, fully consistent with our fiducial requirement.

When local gas densities are sufficiently high, soft X-ray photons absorbed by neutral gas eject electrons from atoms, leading to photoionization. Photons with $E > 54$ eV can ionize He$^{+}$, and subsequent recombinations produce \heii emission lines. These recombination photons have much lower energies than the ionization threshold of neutral hydrogen (13.6 eV, for example, \heiioptic corresponds to 2.65 eV and \heiiUV to 7.56 eV), so they are unaffected by photoelectric absorption by neutral gas. Instead, they suffer only dust attenuation, allowing a significant fraction of \heii emission to escape even when the original soft X-ray continuum is heavily absorbed. This explains why \heii emission can be detected even when the underlying SSS population is often undetected.

Moreover, the intrinsic numbers inferred from nearby galaxies \citep{DiStefano94} should be treated as lower limits, as they exclude other potential soft ionizing sources such as HSSs and QSSs, which had not yet been discerned but may also contribute to He$^+$ ionization. Since our model is agnostic to the physical origin of these sources and instead characterizes them by their spectral energy distribution, the total population of $\sim$3,000 sources inferred above should be interpreted as a combined contribution from classical SSSs and the broader extensions of the class discussed in Section \ref{ssec:extension}, including HSSs, QSSs, and ULSs, all of which emit in the soft X-ray regime and may contribute to He$^+$ ionization. A more complete census of these subpopulations is still needed.

Importantly, the most efficient He$^+$ ionizing temperatures in our models lie around $kT = 20–30$ eV, where the blackbody spectra peak at photon energies just above the He$^+$ ionization threshold. While HSSs (defined as $kT \lesssim 25$ eV) likely contribute, their peak energy output lies just below this optimal range. This suggests that both HSSs and SSSs play a role in \heii ionization, but the dominant contributors must include sources hotter than most HSSs, with spectra extending beyond the far-UV. Future work is needed to assess the relative population sizes and contributions of each class.

For high-redshift NIRSpec galaxies, the required soft ionizing contribution is significantly larger. Our models imply a total soft ionizing luminosity of approximately $3 \times 10^{41}$ to $3 \times 10^{42}$ erg s$^{-1}$ (corresponding to $f_{bb} = 0.1$--1). While SSSs have not been directly observed at high redshift, this luminosity range serves as a benchmark for future studies. The number of sources required to produce this luminosity depends on the currently unknown luminosity distribution of the SSS population, which may vary across cosmic time and environments.

Table~\ref{tab:sss_counts} also shows that star-forming galaxies tend to host more SSSs than those with little star formation. Nearby star-forming spirals (e.g., M83, M101) host significantly more SSSs than passive ellipticals, consistent with the expectation that SSS production is enhanced in young and intermediate-age stellar populations. This provides an intuitive explanation for the larger SSS contributions ($f_{bb}$) inferred in high-redshift galaxies. Star formation rates are systematically higher at $z \sim 2$--6, and these galaxies are expected to host more numerous and younger binary systems, potentially increasing the formation rate of SSSs.

In practice, the required number of SSSs also depends on the true luminosity distribution of the population. While we adopt $10^{38}$ erg s$^{-1}$ as a fiducial value, observed SSSs span a wide range from $\sim10^{37}$ to $10^{39}$ erg s$^{-1}$, with some sources likely extending up to a few times $10^{39}$ erg s$^{-1}$. A more accurate estimate will require incorporating a realistic SSS luminosity function.

In summary, under reasonable assumptions, the required SSS population is consistent with expectations for local star-forming galaxies and plausibly elevated in high-redshift systems. This supports the feasibility of an SSS-driven He$^+$ ionizing contribution across a wide range of galaxy environments.

\subsection{Contributors to the SSS Population}
In our models, the ionizing source responsible for He$^{+}$ ionization is treated phenomenologically, using blackbody spectra with temperatures in the range $kT = 10$–$100$ eV. This spectral regime corresponds to the supersoft X-ray band, which is highly efficient at producing \heii emission via photoionization. We do not assign a specific physical identity to the source of this radiation; rather, we emphasize that any object capable of emitting in this energy range could contribute to the observed \heii flux.

This includes a wide variety of potential contributors: accreting white dwarfs, post-nova systems, stripped stars, X-ray binaries, and even certain phases of WR stars. WR stars, especially in binary systems, emit soft X-rays and may fall within our modeled temperature range. For example, the WR–black hole binary in IC 10 is a confirmed X-ray emitter \citep{Bauer03, Bhattacharya23}, and \cite{Graefener15} demonstrated that WR stars could produce significant He$^+$ ionizing flux, particularly at low metallicities.

Novae also contribute to this picture. Many novae enter a supersoft X-ray phase shortly after eruption, during which they emit significant soft X-ray radiation \citep{Ness07, Ness13}. Optical spectroscopic studies have shown that during this phase, novae can also produce strong \heii emission. For example, KT Eri developed a prominent \heiioptic line that peaked as the nova entered its SSS phase, and persisted through its return to quiescence \citep{Munari14}. Similar behavior has been seen in other novae, including V4743 Sgr and V2491 Cyg, where \heiioptic was observed post-outburst \citep{Zemko18}. These lines are produced as the nova ejecta become optically thin and ionizing photons from the central white dwarf emerge. Although the SSS phase in novae is short-lived, the high occurrence rate of novae in star-forming galaxies makes them a potentially important, transient component of the soft X-ray photon budget.

Given these observational ambiguities and the diversity of potential ionizing sources, our phenomenological modeling provides a flexible framework. By representing all supersoft contributors using blackbody spectra, we quantify the total luminosity necessary to explain observed \heii/\hbeta\ ratios without relying on assumptions about the detailed physical nature of these sources. In future work, we aim to distinguish and quantify the relative contributions of different physical channels explicitly. This approach aligns with our broad definition of SSSs outlined in Section~\ref{sec:SSSs}, where sources are classified based on their temperatures and luminosities rather than specific physical origins. Thus, our model naturally incorporates classical SSSs as well as related populations such as hypersoft sources (HSSs) and ultraluminous SSSs (ULSs), which may become increasingly significant in high-luminosity or high-redshift galaxies. Our framework is sufficiently general to include all such sources, provided they emit radiation energetic enough to ionize He$^+$.

\subsection{Caveats to the Estimated SSS Population}
While our modeling focuses on soft X-ray emitters, additional ionization mechanisms may contribute significantly to the observed \heii budget. Alternative mechanisms such as fast radiative shocks, low-level AGN activity, and non-standard stellar evolutionary pathways have been proposed as potential sources of nebular \heii emission(\citealt{Garnett91, Thuan05, Shirazi12}). Radiative shocks can produce He$^+$-ionizing photons, but only if they are strong enough to provide a substantial fraction of the hydrogen-ionizing flux \citep{Dopita96, Izotov12}. Population III stars may explain \heii emission at $z > 6$ but are unlikely to contribute at lower redshifts where most \heii emitters are observed \citep{GonzalezDiaz25}. Cosmic ray ionization, though theoretically plausible, would require fluxes much higher than those observed in typical star-forming environments \citep{Peimbert72}.

If these alternative mechanisms contribute significantly to He$^+$ ionization, the inferred number of soft X-ray sources in Section~\ref{ssec:estimate} should be regarded as an upper limit. Future work is therefore needed to disentangle the contribution of each ionization mechanism, including SSSs, shocks, Pop III stars, and cosmic rays, to the total He$^+$ ionization budget across redshift and environment.

\begin{figure}
    \centering
    \includegraphics[width=1.0\linewidth]{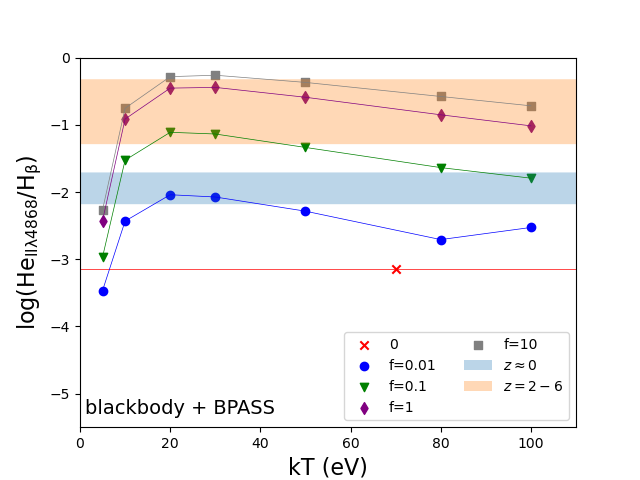}
    \caption{\heii/$\mathrm{H_\beta}$ as a function of blackbody temperature for the MAPPINGS model with a combination of a stellar population spectrum and a blackbody as an input. The bolometric luminosity is kept constant at $L = 10^{39} \mathrm{erg/s}$ and the ionization parameter is kept at $\log U = -3$. Symbols with different colors marks different luminosity fraction of blackbody to the BPASS spectra as indicated in the legend. Red cross marks the MAPPINGS output from the 1 Myr stellar population synthesis from BPASS with a metallicity of $12+\log(\mathrm{O/H})=8.47$ and the gas-phase metallicity is assumed to be similar to the stellar metallicity. The blue shaded region shows the range of the \heii/$\mathrm{H_\beta}$ ratio observed in the SDSS sample and the orange shaded regions shows the observed range in the NIRSpec sample.}
    \label{fig:He-combo_L}
\end{figure}

\begin{figure}
    \centering
    \includegraphics[width=1.0\linewidth]{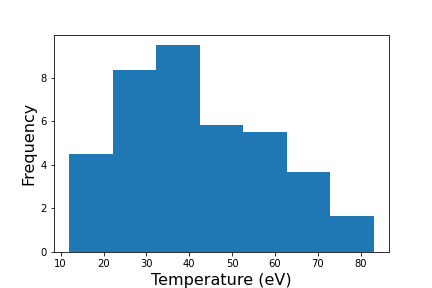}
    \caption{Number of observed supersoft sources as a function of temperature in eV based on \citet{Greiner96} catalog.}
    \label{fig:sss_temperature_histogram}
\end{figure}

\section{Summary} \label{sec:summary}
The origin of \heii lines observed in star-forming galaxies remains a long-standing mystery, as stellar sources alone cannot produce the high-energy photons required to ionize \heii. Previous studies have proposed several alternatives, including ultra lumnious X-ray sources, high-mass X-ray binaries, superbubbles, PopIII stars, radiative shocks, photon leaks, and accreting binaries.
While these mechanisms can produce high-energy photons, they often fail to fully explain the observed \heii/\hbeta ratios in local and high-redshift star-forming galaxies.

We proposed supersoft X-ray sources as the primary contributors to the \heii ionization in star-forming galaxies. SSSs, which can be modeled as blackbodies with temperatures ranging from $kT=10-100$ eV, have been observed in nearby galaxies. However, their low temperature make them difficult to detect, as the soft photons they produced are easily absorbed. Consequently, the observed population of SSSs often under represents the true total number. \citet{DiStefano94} demonstrated that the fraction of observed SSSs decreases with lower temperature higher gas density. 

Using the MAPPINGS photoionization code, we tested the ability of SSSs to reproduce the observed \heiioptic/$\mathrm{H_\beta}$ ratio and explored the parameters that influence \heii ionization. Modeling SSSs as a blackbodies with temperatures between $5-100$ eV, we found in Figure \ref{fig:He-Qvar} that temperature significantly affects the resulting \heii/\hbeta ratio. While we also examined the role of the gas ionization parameter, its impact on the \heii/\hbeta ratio was relatively small. 

To explore the contribution of SSSs to \heii ionization in star-forming galaxies, we combined stellar spectra with blackbody models in the photoionization code (Figure \ref{fig:He-combo_L}). The stellar population SED is adopted from BPASS with an age of 1 Myr and a metallicity of $Z=0.4\ \Zsun$. While keeping the luminosity of the total system constant at $L=10^{39} \mathrm{erg/s}$, we varied the temperature of the blackbody and its luminosity fraction. Model predictions were then compared to observed measurements from local and high-redshift galaxies.

Considering the results of the photoionization model, we conclude that the primary sources of \heii-ionizing photons in local starforming galaxies are SSSs with: $kT=10-30 \mathrm{eV}$ and a low blackbody luminosity fraction ($f_{bb} = 0.01$), $kT=5-10 \mathrm{eV}$ and $kT=100 \mathrm{eV}$ with $f_{bb} = 0.1$. In contrast, the higher \heii/\hbeta ratios observed in high-redshift galaxies (typically about 1 dex above local values) require stronger \heii ionization. These values can be matched by models with $f_{bb} = 0.1$–$10$ and $kT = 10$–$100$ eV.

We also translated our model requirements into physical estimates of SSS populations. Assuming a typical ionizing luminosity of $L_{\mathrm{ion}} = 10^{40}$–$10^{42}$ erg s$^{-1}$ and individual SSS luminosities of $L_{\mathrm{SSS}} = 10^{37}$–$10^{39}$ erg s$^{-1}$, we estimate that between $\sim 30$ and $3 \times 10^4$ SSSs are needed in local galaxies, while $300$ to $3 \times 10^5$ SSSs are required in high-redshift galaxies to reproduce the observed \heii/\hbeta\ ratios. These numbers are consistent with local observations once detection biases and absorption are taken into account, and suggest that soft X-ray sources—while under-observed—could play a major role in shaping the ionizing radiation field in galaxies.

The differences in \heii\ ionization levels between local and high-redshift galaxies likely reflect variations in their underlying SSS populations. Further studies on SSS populations and their host galaxies are needed to investigate potential correlations between the SSS properties and host galaxy properties, and to identify the physical drivers of such trends. Additionally, future work should explore the interplay between SSSs and other ionizing mechanisms—such as radiative shocks, photon leakage, superbubbles, Pop~III stars, and binary-produced processes—to build a more complete picture of the sources of He$^{+}$ ionization across cosmic time. We emphasize that SSSs, whatever their origin, are a natural component of ordinary stellar populations. Some are hot white dwarfs or the central stars of planetary nebulae, while others are binaries containing white dwarfs, neutron stars, or black holes. Ideally, population synthesis models that include binaries should also include SSSs and their cooler counterparts, hypersoft sources (HSSs). This remains a challenge, as we do not yet fully understand the physical nature of many soft sources, and current observations likely capture only a small fraction of the true population. Accounting for this hidden SSS population is crucial to understanding their role in shaping the ionizing radiation field in galaxies.

We also translated our model requirements into physical estimates of SSS populations. Assuming a total bolometric luminosity of $L_{\mathrm{bol}} = 3 \times 10^{43}$ erg s$^{-1}$, a 10\% contribution from a 1~Myr-old stellar population, and individual SSS luminosities of $L_{\mathrm{SSS}} = 10^{37}$--$10^{39}$ erg s$^{-1}$, we estimate that approximately 30 to $3 \times 10^4$ SSSs are needed in local galaxies. For high-redshift galaxies, the required number increases to 300 to $3 \times 10^5$, depending on the assumed SSS luminosity and contribution fraction. These numbers are consistent with local observations once detection biases and absorption are taken into account, and suggest that soft X-ray sources, although under-observed, could play a major role in shaping the ionizing radiation field in galaxies. However, in the most extreme cases, such as when $f_{\mathrm{bb}} = 1$ and $L_{\mathrm{SSS}} = 10^{37}$ erg s$^{-1}$, the required SSS population may exceed the plausible range supported by local analogs. This tension may reflect differences in high-redshift stellar populations or uncertainties in the SSS bolometric corrections, which are often inferred from X-ray observations rather than full spectral energy distributions.

The differences in \heii\ ionization levels between local and high-redshift galaxies likely reflect variations in their underlying SSS populations. Further studies of SSS demographics and their host galaxies are needed to identify potential correlations between SSS properties and intrinsic galaxy characteristics. Additionally, future work should explore the interplay between SSSs and other ionizing mechanisms, such as radiative shocks, photon leakage, superbubbles, Population~III stars, and binary-driven processes, to build a more complete picture of the sources of He$^{+}$ ionization across cosmic time. We emphasize that SSSs, whatever their origin, are part of ordinary stellar populations. Some are hot white dwarfs or the central stars of planetary nebulae, while others are binaries containing white dwarfs, neutron stars, or black holes. Ideally, population synthesis models that include binaries should also include SSSs and their cooler counterparts, hypersoft sources (HSSs). This remains a challenge, as the physical nature of many soft sources is still not well understood, and current observations likely capture only a small fraction of the total population. Accounting for this hidden SSS population is crucial to understanding their role in shaping the ionizing radiation field in galaxies.

%% Also note that the akcnowlodgment environment does not support long amounts of text. If you have a lot of people and institutions to acknowledge, do not use this command. Instead, create a new \section{Acknowledgments}.

\begin{acknowledgments}
DPT would like to thank Yifei Jin and Peixin Zhu for discussion regarding HeII emission in galaxies. (Some of) The data products presented herein were retrieved from the Dawn JWST Archive (DJA). DJA is an initiative of the Cosmic Dawn Center (DAWN), which is funded by the Danish National Research Foundation under grant DNRF140.

\end{acknowledgments}


\vspace{5mm}
%\facilities{HST(STIS), Swift(XRT and UVOT), AAVSO, CTIO:1.3m,


\software{astropy \citep{2013A&A...558A..33A,2018AJ....156..123A},  
          msaexp \citep{msaexp}
          }



%\appendix

%\section{Appendix information}


\bibliography{sample631}{}
\bibliographystyle{aasjournal}

\end{document}